# General Transformations of Space and Time according to Aether Theory

Joseph Levy
4 Square Anatole France, 91250 St Germain lès Corbeil, France
Levy.joseph@orange.fr

Assuming the existence of a fundamental aether frame and the anisotropy of the one-way speed of light in platforms different from the aether frame, we derive the space and time transformations relative to bodies moving in any direction of space and not only in the direction of the common $x$-axis of the co-ordinate systems under consideration. Taking for granted length contraction and clock retardation, we show that the experimental space-time transformations result from measurement distortions due to the fact that the length of the rods and the frequency of the clocks, used for the measurement, do not have a constant value as a result of their motion through the aether, and because the standard synchronization procedures are affected by a synchronism discrepancy effect. When the motion of bodies is aligned along the common $x$-axis, the transformations assume the same mathematical form as the conventional transformations. However, their meaning is quite different because they have been derived on the basis of very different assumptions, and they arise from the measurement distortions mentioned above. Therefore they conceal hidden variables which are the true transformations.

## I. Introduction

It is a common belief that aether theory and special relativity (SR) have the same predictive power, and that the credit given to one theory rather than to the other is a question of philosophical preference [1]. However, the predictive power is not the only interesting thing in science whose role is more specifically to highlight the nature of the physical reality. As an example, it is not indifferent to know whether the one-way speed of light is isotropic in all inertial frames (SR) or not (Eth Th), even if the predictions of the theories which assume these different postulates can be the same.

It is also justified to wonder whether the future predictions will be always the same, and there are good reasons to cast doubt to this. In any cases, according to aether theory, the experimental transformations do not have the same meaning as for SR. They result from unavoidable measurement distortions and conceal hidden variables which are the true transformations. However, even though they need an adjustment, they are useful because we need them to disclose the hidden variables they conceal.

Until now, the transformations of space and time have been focused on motions along the common *x*-axis of two co-ordinate systems, but not along all directions of space [2] The aim of this paper is to derive the transformations according to aether theory in any direction of space and then to compare them to the conventional transformations in specific directions. This would permit to distinguish theoretically the two theories and give criteria for experimental testing.

We specify that the assumptions underlying this study are those defined by Lorentz, that is:

Existence of a fundamental aether frame in which the speed of light is isotropic.
Anisotropy of light speed in all other frames.
Length contraction along the direction of the Earth absolute velocity.
Clock retardation in frames moving relative to the preferred frame.

Our study will comprise two steps:

Initially we will derive the transformations connecting any co-ordinate system moving at constant speed and the preferred frame.

Secondly, we will consider the general case where the transformations connect any two co-ordinate systems, both moving at constant speed with respect to the preferred frame.

## II. Transformations of space and time connecting the aether frame to any co-ordinate system moving at constant speed

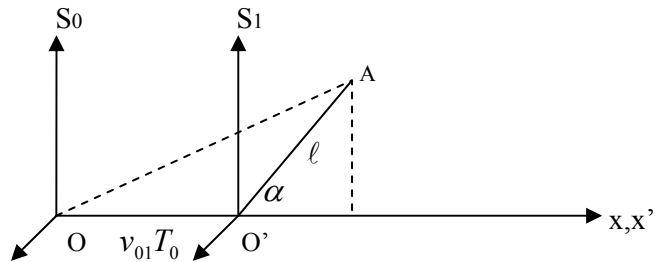

Figure 1. At the initial instant the co-ordinate systems $S_0$ and $S_1$ overlap, at this same instant, a vehicle moving rectilinearly and uniformly, passes by the common origin, before continuing its travel along the rigid path $O'A$. When the vehicle reaches point A, the origin of the system $S_1$ has moved away from $O$, a distance $OO' = v_{01}T_0$.

Let us consider two co-ordinate systems $S_0$ and $S_1$. $S_0$ is at rest in the cosmic substratum (aether frame), and $S_1$ moves at constant speed along the common *x,x'*-axis. At the initial instant the two co-ordinate systems overlap; at this same instant, a vehicle passes by the common origin and heads straight uniformly along a rigid path O'A which makes an angle $\alpha$ with respect to the *x,x'*-axis.

When the vehicle reaches point A, the origin of the system $S_1$ has moved from $O$, a distance $OO' = v_{01}T_0$ (Figure 1).

Our objective is to compare the time and distance needed by the vehicle to reach point A, as measured by an observer at rest relative to the co-ordinate system $S_0$, with the *apparent* time and distance measured in $S_1$.

To measure the *apparent* time, we must synchronize two clocks placed in O' and A. To this end we must beforehand determine the length of the rigid path O'A which has been reduced because of length contraction.

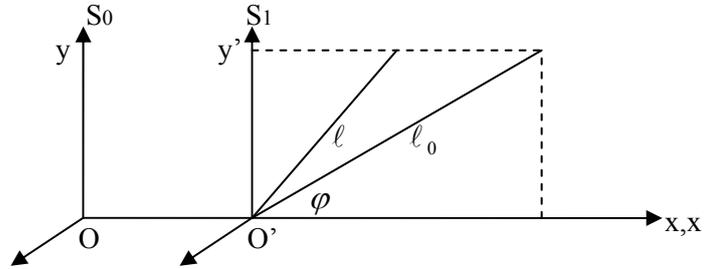

Figure2. Along the *x, x'*-axis, the projection of the rigid path $\ell_0$ contracts, along the y'-axis it is not modified. As a result, the length of the path has been reduced to $\ell$ =O'A, and its direction makes an angle $\alpha$ (not indicated in the figure) with respect to the x,x'-axis.

Let us denote by $\ell_0$ the length assumed by the rigid path when it is at rest in the aether frame, and by $\ell$ its real length in the moving co-ordinate system. We note that along the *x,x'*-axis, the projection of this length is reduced in accordance with the Lorentz Fitzgerald contraction, while along the *y'*-axis it is not modified (Figure 2). Therefore:

$$\ell_0 \cos\varphi = \frac{\ell \cos\alpha}{\sqrt{1 - v_{01}^2/C^2}}$$

and

$$\ell_0 \sin\varphi = \ell \sin\alpha$$

where $\varphi$ is the angle separating $\ell_0$ from the *xx'*-axis, and $\alpha$ the angle between the *xx'*-axis and $\ell$

From the Pythagorean law we have:

$$\left(\frac{\ell\cos\alpha}{\sqrt{1-v_{01}^2/C^2}}\right)^2 + (\ell\sin\alpha)^2 = \ell_0^2.$$

Thus:

$$\ell = \frac{\ell_0\left(1-v_{01}^2/C^2\right)^{1/2}}{\left(1-v_{01}^2\sin^2\alpha/C^2\right)^{1/2}}. \qquad (1)$$

It is important to realize that $\ell$ is the real length of the path in $S_1$ but it is not the measured length in this co-ordinate system. Indeed, the standards used to measure the distance O'A are also contracted in the same ratio, and, therefore, the *apparent* length of O'A in the co-ordinate system $S_1$ is found to be $\ell_0$.

Conversely, an observer in $S_0$ would have obtained an exact estimation of the length O'A because the standards in $S_0$ are not contracted.

**Clock synchronization**

All measurements of time between two distant events carried out in a platform which recedes relative to the aether frame, need beforehand a synchronization of clocks. The methods fluently in use are the Einstein-Poincaré procedure [E-P] with light signals, and the slow clock transport method. These methods have been shown equivalent by different authors [3-11]. Such synchronization procedures could be considered reliable only if the relativity principle was unquestionable and if the speed of light was isotropic. These assumptions are incompatible with the existence of a fundamental aether frame on which aether theory rests [2B]. In this paragraph we will discuss these synchronization procedures on the basis of aether theory.

Assuming the existence of the fundamental aether frame, it can be shown following Prokhovnik [12], that the speed of light in the direction O'A is: (Figure 1)

$$C_{O'A} = -v_{01}\cos\alpha + \sqrt{C^2 - v_{01}^2\sin^2\alpha}, \qquad (2)$$

and from A to O'

$$C_{AO'} = v_{01}\cos\alpha + \sqrt{C^2 - v_{01}^2\sin^2\alpha}.$$

(See the demonstration in appendix 1).

Let us denote by $\tau$ half the two way transit time of light along O'A. We have:

$$\tau = 1/2\,\ell\left(\frac{1}{C_{O'A}} + \frac{1}{C_{AO'}}\right).$$

We easily verify that

$$\tau = \frac{\ell_0}{C\sqrt{1 - v_{01}^2/C^2}}.$$

Taking account of clock retardation, the result of the measurement yields:

$$\tau_{app} = \frac{\ell_0}{C}$$

where the suffix app (for apparent) indicates that, due to clock retardation, the measurement does not provide the exact value of $\tau$.

According to special relativity, $\tau_{app}$ is regarded as the transit time of light along O'A. But insofar as the speed of light is anisotropic, this result is erroneous. The real value being $\dfrac{\ell}{C_{O'A}}$.

In the absence of clock retardation, the difference between the real transit time along O'A and the apparent (measured) time would be:

$$\frac{\ell}{C_{O'A}} - \frac{\ell_0}{C\sqrt{1 - v_{01}^2/C^2}}.$$

But we must take account of clock retardation, and therefore we find:

$$\Delta = \frac{\ell}{C_{O'A}}\sqrt{1 - v_{01}^2/C^2} - \frac{\ell_0}{C}.$$

Where $\Delta$ is the synchronism discrepancy between the clocks placed in O' and A.
From (1) and (2) we obtain:

$$\Delta = \frac{\ell_o[(-v_{01}^2/C)\cos^2\alpha + v_{01}\cos\alpha\sqrt{1 - (v_{01}^2/C^2)\sin^2\alpha}]}{C^2 - v_{01}^2\sin^2\alpha - v_{01}C\cos\alpha\sqrt{1 - (v_{01}^2/C^2)\sin^2\alpha}}.$$

We easily verify that

For $\alpha = 0$ ➔ $\Delta = \dfrac{v_{01}\ell_0}{C^2}$.

We recognize in $\Delta$ the term which appears in the numerator of the transformations relative to time. As we can see, aether theory provides a rational explanation of the meaning of this term, that has no equivalent in special relativity.

For $\alpha = \Pi/2$ ➔ $\Delta = 0$.

### 1. Time transformations for a vehicle moving in any direction of space along a rigid path attached to the moving platform

The real transit time is the time $T_0$ measured with a clock which is not slowed down by motion (i.e; the time that a clock standing in the aether frame would display). Due to clock retardation, the apparent time displayed by a clock placed in O' is: (Figure 1)

$$T_0\sqrt{1-v_{01}^2/C^2}.$$

Because of the synchronism discrepancy, the clock placed in A will display the reading:

$$T_{app} = T_0\sqrt{1-v_{01}^2/C^2} - \Delta$$

Therefore:

$$T_{app} = T_0\sqrt{1-v_{01}^2 C^2} - \frac{\ell_o[v_{01}\cos\alpha\sqrt{1-(v_{01}^2/C^2)\sin^2\alpha} - (v_{01}^2/C)\cos^2\alpha]}{C^2 - v_{01}^2\sin^2\alpha - v_{01}C\cos\alpha\sqrt{1-(v_{01}^2/C^2)\sin^2\alpha}},$$

and:

$$T_0 = \frac{T_{app}}{\sqrt{1-v_{01}^2/C^2}} + \frac{\ell_o[v_{01}\cos\alpha\sqrt{1-(v_{01}^2/C^2)\sin^2\alpha} - (v_{01}^2/C)\cos^2\alpha]}{\sqrt{1-v_{01}^2/C^2}(C^2 - v_{01}^2\sin^2\alpha - v_{01}C\cos\alpha\sqrt{1-(v_{01}^2/C^2)\sin^2\alpha})}$$

From now on, let us examine the special cases where $\alpha = 0$ and $\alpha = \Pi/2$.

**a. Form of the transformations when $\alpha = 0$.**

The expression of $T_{app}$ reduces to:

$$T_{app} = T_0\sqrt{1-v_{01}^2/C^2} - \frac{[v_{01}\ell_0 - v_{01}^2\ell_0/C]}{C^2 - v_{01}C},$$

$$T_{app} = T_0\sqrt{1-v_{01}^2/C^2} - \frac{v_{01}\ell_0}{C^2}.$$

Noting that the apparent distance from O' to A, (as measured with a contracted standard of the moving platform $S_1$) is equal to $\ell_0$, we will denote $\ell_0$ by $X_{app}$. Therefore:

$$T_{app} = T_0\sqrt{1-v_{01}^2/C^2} - \frac{v_{01}X_{app}}{C^2},$$

and:

$$T_0 = \frac{T_{app} + v_{01}X_{app}/C^2}{\sqrt{1-v_{01}^2/C^2}}. \tag{3}$$

This expression has the same mathematical form as the conventional transformation relative to time, but its meaning is quite different because $X_{app}$ and $T_{app}$ are shown to be the apparent co-ordinates of the moving platform, and because $C$ is not the speed of light in all inertial frames, it is the speed of light in the aether frame exclusively.

**b. For $\alpha = \Pi/2$ we have:**

$$T_{app} = T_0 \sqrt{1 - v_{01}^2/C^2},$$

and

$$T_0 = \frac{T_{app}}{\sqrt{1 - v_{01}^2/C^2}}. \qquad (4)$$

Expressions (3) and (4) highlight the role of the aether drift. The term $\dfrac{v_{01} X_{app}}{C^2}$ which translates the synchronism discrepancy effect does not exist when the measurement is made in a direction perpendicular to the drift. Special relativity has no rational explanation for this result. Actually, for special relativity, given that there is no preferred aether frame, $T_0$ does not mean anything and the measured time along the direction $\alpha = \Pi/2$ is regarded as the real transit time of the vehicle.

## 2. Space transformations for a vehicle moving in any direction of space along a rigid path attached to the moving platform

Relative to the moving platform, the path covered by the vehicle is O'A, but relative to the aether frame, it is OA. We will therefore denote OA by $X_0$ (Figure 1).

Given that the real length of O'A is $\ell$, we have:

$$X_0^2 = (v_{01} T_0 + \ell \cos\alpha)^2 + \ell^2 \sin^2\alpha.$$

Therefore:

$$X_0 = \sqrt{(v_{01} T_0 + \ell \cos\alpha)^2 + \ell^2 \sin^2\alpha}$$

$$= \sqrt{v_{01}^2 T_0^2 + 2\ell v_{01} T_0 \cos\alpha + \ell^2}$$

Replacing $\ell$ by its value given in (1) yields:

$$X_0 = \sqrt{\frac{\ell_0^2 (1 - v_{01}^2/C^2)}{1 - v_{01}^2 \sin^2\alpha/C^2} + v_{01}^2 T_0^2 + 2\ell_0 v_{01} T_0 \cos\alpha \frac{(1 - v_{01}^2/C^2)^{1/2}}{(1 - v_{01}^2 \sin^2\alpha/C^2)^{1/2}}}.$$

From now on, let us examine the special cases where $\alpha = 0$ and $\alpha = \Pi/2$.

**a. Form of the transformations when $\alpha = 0$.**

The expression of $X_0$ reduces to:

$$X_0 = \sqrt{\ell_0^2(1 - v_{01}^2/C^2) + v_{01}^2 T_0^2 + 2v_{01}\ell_0 T_0 (1 - v_{01}^2/C^2)^{1/2}}$$
$$= v_{01}T_0 + \ell_0\sqrt{1 - v_{01}^2/C^2}.$$

$\ell_0$ is the real value of the rigid path when it is at rest in the co-ordinate system $S_0$, but it is also the apparent value $X_{app}$ in $S_1$, as measured with a contracted standard.

Replacing $\ell_0$ by its value $X_{app}$, and $T_0$ by the expression $\dfrac{T_{app} + v_{01}X_{app}/C^2}{\sqrt{1 - v_{01}^2/C^2}}$ derived in the previous paragraph yields:

$$X_0 = \frac{X_{app}(1 - v_{01}^2/C^2) + v_{01}(T_{app} + v_{01}X_{app}/C^2)}{\sqrt{1 - v_{01}^2/C^2}}.$$

Hence

$$X_0 = \frac{X_{app} + v_{01}T_{app}}{\sqrt{1 - v_{01}^2/C^2}}. \tag{5}$$

The reciprocal transformations for $\alpha = 0$ can be easily derived from (3) and (5). They are:

$$T_{app} = \frac{T_0 - v_{01}X_0/C^2}{\sqrt{1 - v_{01}^2/C^2}} \tag{6}$$

and

$$X_{app} = \frac{X_0 - v_{01}T_0}{\sqrt{1 - v_{01}^2/C^2}}. \tag{7}$$

Although these transformations take the same mathematical form as the conventional transformations, they differ from them in many respects:

1. $X_{app}$ and $T_{app}$ result from measurement distortions because they are measured with contracted standards and clocks slowed down by motion and arbitrarily synchronized. Therefore, the transformations conceal hidden variables which are the true transformations. (Note that, the similarity with the conventional transformations is only apparent. Indeed, for special relativity, $X_0$ and $T_0$ have no meaning, since there is no aether frame).

(The true transformations are the Galilean transformations. Of course, the increase of mass with speed implies that the speed of a body ($V$) relative to the aether frame must be limited in such a way that $V<C$. This means that when a body A moves at speed $v_A$ from the origin of a co-ordinate system which is at rest with respect to the aether frame, the speed relative to A of another body B moving along the direction OA will be limited to $v_B < C - v_A$) [2F].

2. The transformations have not been derived on the assumption of the relativity principle and of the invariance of the speed of light. Therefore, the complete symmetry of the transformations, which is the cause of most of the difficulties encountered by SR, no longer exists, as can be seen in expressions (3) and (5) and in (6) and (7).

**b. For $\alpha = \Pi/2$ (noting that in this direction $\ell_0 = \ell$) we have:**

$$X_0 = \sqrt{\ell_0^2 + v_{01}^2 T_0^2} = X_{app}\sqrt{1+(\frac{v_{01}}{v_{O'A}})^2} \ . \tag{8}$$

This result has no equivalence in special relativity for which the system $S_0$ does not exist.

### III. Transformations of space and time connecting any pair of co-ordinate systems receding from one another at constant speed.

Let us consider three co-ordinate systems $S_0$, $S_1$ and $S_2$. $S_0$ is at rest in the cosmic substratum (aether frame), while $S_1$ and $S_2$ move uniformly along the common *x*-axis. At the initial instant, the three co-ordinate systems overlap. At this instant a vehicle passes by the common origin and heads straight uniformly along a rigid path O''A which is firmly tied to the co-ordinate system $S_2$ (Figure 3).

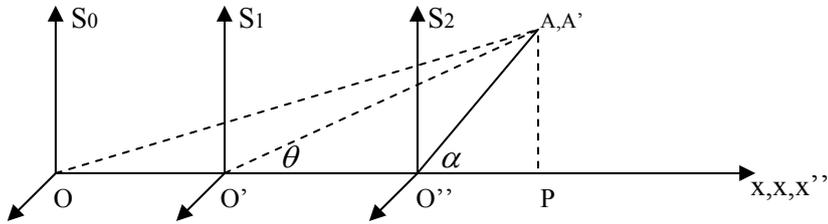

Figure3. At the initial instant, the three co-ordinate systems overlap, at this instant, a vehicle, passing by the common origin, moves at constant speed along a rigid path O''A firmly tied to the co-ordinate system $S_2$.

With respect to the co-ordinate system $S_2$ the path covered by the vehicle is O''A, but with respect to $S_1$ it is O'A', where A' is a fixed point of the co-ordinate $S_1$ which coincides with point A when the vehicle reaches this point.

Our objective is **1**. To compare the *apparent* distances covered by the vehicle in order to reach AA' when measurements are made with the contracted standards of $S_1$ and $S_2$. From these relative distances we will be able to show that the experimental results follow from the systematic distortions which affect the Galilean co-ordinates when standard measurements are used.

**2.** To compare the apparent times needed by the vehicle to reach point AA' as measured with clocks of the systems $S_1$ and $S_2$ which are slowed down by motion and are synchronized by the Einstein-Poincaré procedure (E-P), and to show that the experimental time transformations result from the systematic distortions which affect the measurement of the universal time.

To this end, we start from the Galilean co-ordinates, and we apply to them the distortions which are systematic and cannot be avoided when measurements are made, showing that the experimental space-time transformations conceal hidden variables which are the Galilean transformations.

1. **Space transformations for a vehicle travelling in any direction of space along a rigid path attached to the coordinate system $S_2$**

The ratio of the real distances covered by the vehicle with respect to O' and O'' is equal to the ratio of the real speeds relative to these points.
Therefore:

$$\frac{O'A'}{O''A} = \frac{v_{O'A'}}{v_{O''A}}.$$

Denoting O'A' by $X_{1r}$ we have:

$$X_{1r} = \frac{v_{O'A'}}{v_{O''A}} O''A,$$

where the suffix r is put for real.

From the Pythagorean relation, we can express $v_{O'A'}$ as a function of $v_{O'P}$ and $v_{PA'}$ (Figure 3).

As for the length of O''A, it is determined from the relation (1).

Therefore:

$$X_{1r} = \ell_O \frac{\sqrt{(v_{12} + v_{O''A}\cos\alpha)^2 + (v_{O''A})^2 \sin^2\alpha}}{v_{O''A}} \times \frac{\sqrt{1 - v_{02}^2/C^2}}{\sqrt{1 - v_{02}^2 \sin^2\alpha/C^2}}. \quad (9)$$

Now, the measurement of O'P by an observer attached to the co-ordinate system $S_1$ is made with standards whose length is contracted because of the motion of this co-ordinate system with respect to the aether frame.

The apparent length of O'P is therefore equal to:

$$\frac{X_{1r} \cos\theta}{\sqrt{1 - v_{01}^2/C^2}}.$$

Thus, with his instruments, an observer attached to $S_1$ will find for $(O'A')^2$ a value equal to:

$$X_{1app}^2 = \left(\frac{X_{1r}\cos\theta}{\sqrt{1 - v_{01}^2/C^2}}\right)^2 + X_{1r}^2 \sin^2\theta,$$

therefore:

$$X_{1app} = X_{1r}\sqrt{\frac{\cos^2\theta}{1 - v_{01}^2/C^2} + \sin^2\theta}$$

$$= \frac{X_{1r}\sqrt{1 - v_{01}^2 \sin^2\theta/C^2}}{\sqrt{1 - v_{01}^2/C^2}}. \quad (10)$$

From (9) and (10) given that $X_{2app} = \ell_0$, we obtain:

$$X_{1app} = X_{2app} \frac{\sqrt{(v_{12} + v_{O"A}\cos\alpha)^2 + (v_{O"A})^2 \sin^2\alpha}}{v_{O"A}} \times \frac{\sqrt{1 - v_{02}^2/C^2}}{\sqrt{1 - v_{02}^2 \sin^2\alpha/C^2}} \times \frac{\sqrt{1 - v_{01}^2 \sin^2\theta/C^2}}{\sqrt{1 - v_{01}^2/C^2}}$$

(11)

From now on, let us examine the special cases where $\alpha = 0$ and $\alpha = \Pi/2$.

**a. Form of the transformations when $\alpha = 0$.**

For $\alpha = 0$ (and $\theta = 0$), expression (11) reduces to:

$$X_{1app} = X_{2app} \frac{(v_{12} + v_{O"A})}{v_{O"A}} \frac{\sqrt{1 - v_{02}^2/C^2}}{\sqrt{1 - v_{01}^2/C^2}}.$$

Denoting the speed of the vehicle with respect to $S_0$ by $V$, this expression can be written as:

$$X_{1app} = X_{2app} \frac{(V - v_{O1})}{(V - v_{O2})} \frac{\sqrt{1 - v_{02}^2/C^2}}{\sqrt{1 - v_{01}^2/C^2}} \tag{12}$$

which is the expression we obtained in ref [2].

After some calculations, expression (12) takes the form:

$$X_{1app} = \frac{X_{2app} + \dfrac{v_{02} - v_{01}}{1 - v_{01}v_{02}/C^2} T_{2app}}{\sqrt{1 - \dfrac{(v_{02} - v_{01})^2}{C^2(1 - v_{01}v_{02}/C^2)^2}}}. \tag{13}$$

(See appendix 2 for the demonstration).

Denoting $\dfrac{v_{02} - v_{01}}{1 - v_{01}v_{02}/C^2}$ by $v_{12app}$, expression (13) can be written as:

$$X_{1app} = \frac{X_{2app} + v_{12app} T_{2app}}{\sqrt{1 - v_{12app}^2/C^2}}. \tag{14}$$

Expression (14) takes the same mathematical form as the conventional transformation relative to space, but obviously its meaning is quite different. It shows that this mathematical form permits to predict the measured experimental data but not the true ones, which, due to measurement distortions, are unapparent. A correction to expression (14) is therefore necessary to highlight the hidden variables which are the Galilean transformations.

**b. For $\alpha = \Pi/2$, we have:**

$$X_{1app} = X_{2app} \frac{\sqrt{v_{12}^2 + v_{O''A}^2}}{v_{O''A}} \sqrt{\frac{\cos^2 \theta}{1 - v_{01}^2/C^2} + \sin^2 \theta}$$

$$X_{1app} = X_{2app} \frac{v_{O'A'}}{v_{O''A}} \sqrt{\frac{1 - v_{01}^2 \sin^2 \theta / C^2}{1 - v_{01}^2/C^2}}$$

## 2. Time transformations for a vehicle travelling in any direction of space along a rigid path attached to the coordinate system S$_2$

As a starting point, we assume that the true time is the same in all moving platforms. As we shall show, the experimental time transformations translate in fact the distortions which affect the measurement because of the slowing down of clocks moving through the aether, and those which result from the synchronization procedures.

Let us denote by $T_r$ the real time needed by the aforementioned vehicle to move from O'' to A. We have:

$$T_r = \ell / v_{O''A},$$

where $\ell$ is the real length of the rigid path from O'' to A which, according to formula (1), is equal to:

$$\ell = \frac{\ell_0 \left(1 - v_{02}^2/C^2\right)^{1/2}}{\left(1 - v_{02}^2 \sin^2 \alpha / C^2\right)^{1/2}}.$$

$v_{O''A}$ is the real speed of the vehicle relative to the co-ordinate system $S_2$.

$T_r$ is the same in all co-ordinate systems, but in the co-ordinate system $S_1$, due to clock retardation, a clock placed in point O' would display the reading:

$$T_r \sqrt{1 - v_{01}^2/C^2}.$$

**Clock synchronization**

In addition, in order to know the *apparent* time displayed by the clocks in $S_1$, that the vehicle takes to reach point AA', we must beforehand synchronize the clocks placed in O' and A'.

According to the E-P procedure, the time needed by a light signal to run from O' to A' is considered equal to half the two way transit time of light

$$\frac{(t_1 + \bar{t}_1)}{2}$$

where $\bar{t}_1$ is the transit time of light from A' to O'.

In reality the true time needed by the light signal to move from O' to A' is equal to $t_1$.

Taking account of clock retardation, the synchronism discrepancy between the clocks placed in O' and A' is:

$$\Delta = t_1\sqrt{1-v_{01}^2/C^2} - \frac{(t_1 + \bar{t}_1)}{2}\sqrt{1-v_{01}^2/C^2} = \frac{(t_1 - \bar{t}_1)}{2}\sqrt{1-v_{01}^2/C^2}.$$

Denoting the speed of light in the two reverse directions by $C_{O'A'}$ and $C_{A'O'}$ we have:

$$C_{O'A'} = -v_{01}\cos\theta + \sqrt{C^2 - v_{01}^2 \sin^2\theta}$$

and

$$C_{A'O'} = v_{01}\cos\theta + \sqrt{C^2 - v_{01}^2 \sin^2\theta}.$$

(See the demonstration in Appendix 1).

Therefore:

$$\Delta = \frac{X_{1r}}{2}\left(\frac{1}{C_{O'A'}} - \frac{1}{C_{A'O'}}\right)\sqrt{1-v_{01}^2/C^2}$$

$$= \frac{1}{2}\left(\frac{2v_{01}\cos\theta}{C^2 - v_{01}^2}\right)X_{1r}\sqrt{1-v_{01}^2/C^2}.$$

Replacing $X_{1r}$ by its value given in formula (9) yields:

$$\Delta = \ell_O \frac{v_{01}\cos\theta}{C^2 - v_{01}^2} \frac{\sqrt{(v_{12} + v_{O''A}\cos\alpha)^2 + (v_{O''A})^2 \sin^2\alpha}}{v_{O''A}} x \frac{\sqrt{1-v_{02}^2/C^2}}{\sqrt{1-v_{02}^2 \sin^2\alpha/C^2}}\sqrt{1-v_{01}^2/C^2}$$

(15)

**Time transformations**

The *apparent* transit time taken by the vehicle to reach point AA' measured with clocks of the co-ordinate system $S_1$, will be equal to:

$$T_{1app} = T_{1r}\sqrt{1-v_{01}^2/C^2} - \Delta \qquad (16)$$

where $T_{1r} = T_{2r} = T_r$

From formula (1) we have:

$$T_{1r} = \frac{\ell}{v_{O''A}} = \frac{\ell_0(1-v_{02}^2/C^2)^{1/2}}{v_{O''A}(1-v_{02}^2 \sin^2\alpha/C^2)^{1/2}}. \qquad (17)$$

From formulas (15), (16) and (17) we obtain:

$$T_{1app} = \ell_0 \frac{\sqrt{1-v_{02}^2/C^2}\sqrt{1-v_{01}^2/C^2}}{v_{O"A}\sqrt{1-v_{02}^2\sin^2\alpha/C^2}}(1-\frac{v_{01}\cos\theta}{C^2-v_{01}^2}\sqrt{v_{12}^2+2v_{12}v_{O"A}\cos\alpha+v_{O"A}^2})$$

. From now on, let us examine the special cases where $\alpha = 0$ and $\alpha = \Pi/2$.

**a. Form of the transformations when $\alpha = 0$.**

For $\alpha = 0$ (and $\theta = 0$), the expression of $T_{1app}$ reduces to:

$$T_{1app} = \ell_0 \frac{\sqrt{1-v_{02}^2/C^2}\sqrt{1-v_{01}^2/C^2}}{v_{O"A}}[1-\frac{v_{01}}{C^2-v_{01}^2}(v_{12}+v_{O"A})].$$

Given that the apparent length $X_{2app}$ of the rigid path O''A measured with a contracted standard is equal to $\ell_0$ we obtain:

$$T_{1app} = X_{2app}\frac{\sqrt{1-v_{02}^2/C^2}}{\sqrt{1-v_{01}^2/C^2}}\frac{(C^2-v_{01}^2)}{C^2}[\frac{C^2-v_{01}^2-v_{01}(v_{12}+v_{O"A})}{v_{O"A}(C^2-v_{01}^2)}].$$

Replacing $v_{12}+v_{O"A}$ by its value $V-v_{01}$ where $V$ is the real speed of the vehicle relative to the aether frame, and replacing $v_{O"A}$ by $V-v_{02}$, yields:

$$T_{1app} = X_{2app}\frac{\sqrt{1-v_{02}^2/C^2}}{\sqrt{1-v_{01}^2/C^2}}(\frac{1-v_{01}V/C^2}{V-v_{02}}). \tag{18}$$

(Note that in aether theory, real speeds obey the Galilean law of composition of velocities. As we shall see, only apparent speeds obey the relativistic law).

From the expressions (6) and (7) we can verify that:

$$\frac{X_{2app}}{T_{2app}} = \frac{V-v_{02}}{1-v_{02}V/C^2}.$$

Replacing $X_{2app}$ with its value in (18) yields:

$$T_{1app} = T_{2app}\frac{\sqrt{1-v_{02}^2/C^2}}{\sqrt{1-v_{01}^2/C^2}}(\frac{1-v_{01}V/C^2}{1-v_{02}V/C^2}), \tag{19}$$

which is the expression we obtained in ref [2].

After some calculations, expression (19) can be written in the form:

$$T_{1app} = \frac{T_{2app} + \dfrac{v_{02} - v_{01}}{1 - v_{01}v_{02}/C^2} \dfrac{X_{2app}}{C^2}}{\sqrt{1 - \dfrac{1}{C^2}(\dfrac{v_{02} - v_{01}}{1 - v_{01}v_{02}/C^2})^2}}$$

(See appendix 2 for the demonstration).

Denoting $\dfrac{v_{02} - v_{01}}{1 - v_{01}v_{02}/C^2}$ by $v_{12app}$, yields:

$$T_{1app} = \frac{T_{2app} + v_{12app} X_{2app}/C^2}{\sqrt{1 - v_{12app}^2/C^2}}. \tag{20}$$

Expression (20) assumes the same mathematical form as the conventional transformation relative to time, but its meaning is quite different; it permits to predict the measured experimental data but not the true ones which, due to measurement distortions, are unapparent. A correction to expression (20) is therefore necessary to highlight the hidden variables which are the Galilean transformations.

**b**. For $\alpha = \Pi/2$, we have:

$$T_{1app} = \frac{\ell_0}{v_{O''A}} \sqrt{1 - v_{01}^2/C^2}(1 - \frac{v_{01} \cos\theta}{C^2 - v_{01}^2}\sqrt{v_{12}^2 + v_{O''A}^2})$$

$$= \frac{\ell_0}{v_{O''A}} \sqrt{1 - v_{01}^2/C^2}(1 - \frac{v_{01} v_{O'A'} \cos\theta}{C^2 - v_{01}^2})$$

$$= \frac{\ell_0}{v_{O''A}} \sqrt{1 - v_{01}^2/C^2}(1 - \frac{v_{01} v_{12}}{C^2 - v_{01}^2}) \tag{21}$$

Expression (21) can also be expressed as a function of $T_{2app}$.

Noting that for $\alpha = \Pi/2$, $\ell = \ell_0$, we have (from expression (4)):

$$T_{2app} = T_r \sqrt{1 - v_{02}^2/C^2}$$

$$= \frac{\ell_0}{v_{O''A}} \sqrt{1 - v_{02}^2/C^2},$$

we obtain:

$$T_{1app} = T_{2app} \frac{\sqrt{1 - v_{01}^2/C^2}}{\sqrt{1 - v_{02}^2/C^2}}(1 - \frac{v_{01} v_{12}}{C^2 - v_{01}^2}).$$

**Note**

When $v_{01} = 0$, the co-ordinate system $S_1$ coincides with $S_0$, so that $T_{1app}$ is reduced to $T_r$. Therefore, (as expected):

$$T_r = \frac{T_{2app}}{\sqrt{1 - v_{02}^2 / C^2}}.$$

## Appendix 1
## Speed of light in any direction of space

Let us consider two co-ordinate systems, $S_0$ and $S$. $S_0$ is at rest in the cosmic substratum (aether frame) and $S$ is attached to a platform which moves with rectilinear uniform motion at speed $v$ along the $x_0$-axis of the system $S_0$ and suppose that a rod MN, making an angle $\alpha$ with the $x_0$, $x$-axis, is at rest with respect to the system $S$ [12, 2G].

At the two ends of the rod, let us place two mirrors facing one another by their reflecting surface, which is perpendicular to the axis of the rod $\ell = MN$. At the initial instant, the two systems $S_0$ and $S$ overlap. At this very instant a light signal is sent from the common origin and travels along the rod towards point N. When the signal reaches this point the rod has been translated to a distance equal to $vt$ and is referred to as *M'N'* where $t$ is the time needed by the signal to cover the distance *MN* (Figure 4).

After reflection the signal reverses its travel. (Note that the length of the moving rod is contracted according to formula (1)).

We remark that the path of the light signal along the rod is related to the speed $C_1$ by the relation:

$$C_1 = \frac{MN}{t}.$$

In addition, when the signal reaches point N, the system $S$ has moved away from $S_0$ a distance *MM'=vt*, so that:

$$v = \frac{MM'}{t}.$$

The same distance has been covered by point N which is translated to N'

Now, from the point of view of an observer which is supposed at rest in $S_0$, the signal goes from point M to point N'(Figure 4).

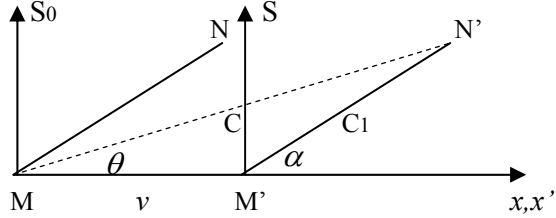

Figure 4. The speed of light is equal to $C_1$ from M' to N' and to C from M to N'.

$C$ being the speed of light in $S_0$, we have:

$$\frac{MN'}{t} = C$$

and hence, the projection along the $x,x'$-axis of the speed of light $C_1$ relative to the system $S$, will be equal to $(C\cos\theta - v)$. So that:

$$C\cos\theta - v = C_1 \cos\alpha .$$

The three speeds, $C$, $C_1$ and $v$ being proportional to the three lengths MN', MN and MM' with the same coefficient of proportionality, we have

$$C^2 = (C_1 \cos\alpha + v)^2 + C_1^2 \sin^2\alpha .$$

Therefore:

$$C_1^2 + 2vC_1 \cos\alpha - (C^2 - v^2) = 0 . \qquad (22)$$

(We must emphasize that equation (22) implies that the three speeds $C$, $C_1$ and $v$ have been measured with the help of the same clock, which obviously is a clock whose clock rate is not slowed down by motion.)

Resolving the second degree equation, yields:

$$C_1 = -v\cos\alpha \pm \sqrt{C^2 - v^2 \sin^2\alpha} .$$

The condition $C_1 = C$ when $v = 0$ compels us to only retain the + sign so:

$$C_1 = -v\cos\alpha + \sqrt{C^2 - v^2 \sin^2\alpha} . \qquad (23)$$

QED

Now, the return of light can be illustrated by the figure 5 below:

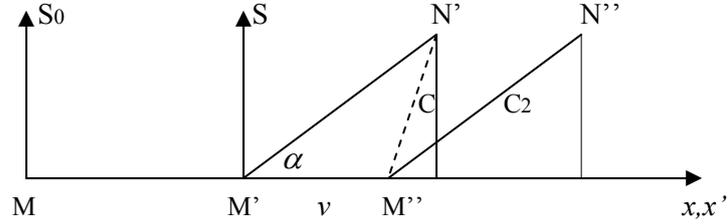

Figure 5. The speed of light is equal to C2 from N'' to M'' and to C from N' to M''. $\theta$' is the angle separating N'M'' from the *x,x'*-axis. (not indicated in the figure)

From the point of view of an observer attached to the system $S$, the light comes back to its initial position with the speed $C_2$.

Therefore we can write:

$$C_2 = \frac{N''M''}{t'}.$$

Where *t'* is the time of light transit from its final to its initial position.

For an observer which is supposed at rest relative to the system $S_0$, the light comes from N' to M'' with the speed *C*, so that:

$$C = \frac{N'M''}{t'}.$$

During the light transfer, the system $S$ has moved from M' to M'' with the speed *v* therefore:

$$v = \frac{M'M''}{t'}.$$

The projection of the speed of light relative to $S$ along the *x,x'*-axis will be:

$$C \cos\theta' + v = C_2 \cos\alpha$$

where $\theta$' is the angle separating N'M'' from the *x,x'*-axis.

We easily verify that:

$$(C_2 \cos\alpha - v)^2 + (C_2 \sin\alpha)^2 = C^2,$$

therefore,

$$C_2 = v\cos\alpha + \sqrt{C^2 - v^2 \sin^2\alpha}. \tag{24}$$

QED

## Appendix 2

Starting from the expressions (12) and (19) we will derive the expressions (14) and (20) whose mathematical form is similar to that of the conventional transformations [2B].

**Space transformations:**

From expression (12), we have successively:

$$X_{1app} = X_{2app} \frac{\sqrt{1-v_{02}^2/C^2}}{\sqrt{1-v_{01}^2/C^2}} \frac{(V-v_{01})}{(V-v_{02})}$$

$$= X_{2app} \frac{(1-v_{02}^2/C^2)(V-v_{01})}{\sqrt{(1-v_{02}^2/C^2)(1-v_{01}^2/C^2)}(V-v_{02})}$$

$$= X_{2app} \frac{V - v_{01} - v_{02}^2 V/C^2 + v_{01}v_{02}^2/C^2}{(V-v_{02})\sqrt{1 - v_{01}^2/C^2 - v_{02}^2/C^2 + v_{01}^2 v_{02}^2/C^4}}$$

$$= X_{2app} \frac{(V-v_{02})(1-v_{01}v_{02}/C^2) + (v_{02}-v_{01})(1-v_{02}V/C^2)}{(V-v_{02})\sqrt{(1-v_{01}v_{02}/C^2)^2 - \frac{(v_{02}-v_{01})^2}{C^2}}}$$

$$= X_{2app} \frac{\dfrac{(V-v_{02})(1-v_{01}v_{02}/C^2) + (v_{02}-v_{01})(1-v_{02}V/C^2)}{(V-v_{02})(1-v_{01}v_{02}/C^2)}}{\sqrt{\dfrac{C^2(1-v_{01}v_{02}/C^2)^2 - (v_{02}-v_{01})^2}{C^2(1-v_{01}v_{02}/C^2)^2}}}$$

$$= \frac{X_{2app} + \dfrac{v_{02}-v_{01}}{1-v_{01}v_{02}/C^2} \dfrac{X_{2app}}{V-v_{02}} \dfrac{1}{1-v_{02}V/C^2}}{\sqrt{1 - \dfrac{(v_{02}-v_{01})^2}{C^2(1-v_{01}v_{02}/C^2)^2}}}.$$

Finally:

$$X_{1app} = \frac{X_{2app} + \dfrac{v_{02}-v_{01}}{1-v_{01}v_{02}/C^2} T_{2app}}{\sqrt{1 - \dfrac{(v_{02}-v_{01})^2}{C^2(1-v_{01}v_{02}/C^2)^2}}}. \qquad (25)$$

Unlike conventional transformations, these expressions highlight the fact that the space and time co-ordinates $X_{app}$ and $T_{app}$ result from systematic measurement errors. Of course these errors cannot be avoided because the measurements are made with standards whose length vary with speed, and with clocks whose frequency changes when they are transferred from one platform to another of different absolute speed. Besides, they suffer from the synchronism discrepancy effect. (Therefore, only a theoretical treatment can restore the true values).

These transformations are not based on the assumption of the indisputable value of the relativity principle, and on the invariance of the speed of light. *C* is the speed in the aether frame and not in all 'inertial' frames. The derivation has been carried out from the Galilean transformations which were subjected to the measurement distortions. Therefore these transformations conceal hidden variables which are the Galilean transformations. They have not been derived from the assumption that they constitute a group, since such an assumption confers the same status to all transformations; yet, here, when the relative speed between $S_1$ and $S_0$ is reduced to zero, these transformations reduce to:

$$X_0 = \frac{X_{2app} + v_{02}T_{2app}}{\sqrt{1 - v_{02}^2/C^2}}.$$

And the reciprocal transformations take the form:

$$X_{2app} = \frac{X_0 - v_{02}T_0}{\sqrt{1 - v_{02}^2/C^2}}.$$

These transformations differ somewhat from the expression (25) and assume a different status since $v_{02}$ is the real speed and differs from the apparent speed $\frac{v_{02} - v_{01}}{1 - v_{01}v_{02}/C^2}$ which appears in expression (25), a result which highlights the fact that the laws of nature are affected by the existence of the aether drift.

Note also that these transformations do not imply the relativity of simultaneity which is a concept quite irrational. For aether theory, the relativity of simultaneity is an artefact depending on the measurement distortions.

Note that, when the speeds are measured with contracted standards and with clocks slowed down by motion and synchronized with light signals, their apparent value is equal to $v_{12app}$ such that :

$$v_{12app} = \frac{v_{02} - v_{01}}{1 - v_{01}v_{02}/C^2}. \tag{26}$$

With these apparent speeds, the space transformations take *the same mathematical form* as the conventional transformations between any pair of 'inertial frames', their general form being:

$$X_{1app} = \frac{X_{2app} + v_{12app}T_{2app}}{\sqrt{1 - v_{12app}^2/C^2}} \quad \text{and} \quad X_{2app} = \frac{X_{1app} - v_{12app}T_{1app}}{\sqrt{1 - v_{12app}^2/C^2}}$$

and therefore the relativity principle *seems* to apply. Actually it does not really strictly apply when measurements are exact, and therefore it gives a distorted view of reality. Moreover, $v_{12app}$ has not the same meaning as in conventional relativity since it depends on the speeds $v_{01}$ and $v_{02}$ of reference frames S$_1$ and S$_2$ with respect to the aether frame. This dependence highlights the importance of the aether in this derivation while in the conventional approaches the aether is hidden or nonexistent. In Einstein's approach $v_{01}$ and $v_{02}$ do not mean anything and in Poincaré's approach the aether frame has no special status compared with the other 'inertial' frames.

**Time transformations:**

We start from expression (19). We have successively:

$$T_{1app} = T_{2app} \frac{\sqrt{1 - v_{02}^2/C^2}}{\sqrt{1 - v_{01}^2/C^2}} \frac{(1 - v_{01}V/C^2)}{(1 - v_{02}V/C^2)}$$

$$= T_{2app} \frac{(1 - v_{02}^2/C^2)(1 - v_{01}V/C^2)}{\sqrt{(1 - v_{01}^2/C^2)(1 - v_{02}^2/C^2)}(1 - v_{02}V/C^2)}$$

$$= T_{2app} \frac{(1 - v_{02}^2/C^2 - v_{01}V/C^2 + v_{01}v_{02}^2V/C^4)}{(1 - v_{02}V/C^2)\sqrt{(1 - v_{01}^2/C^2 - v_{02}^2/C^2 + v_{01}^2v_{02}^2/C^4}}$$

$$= T_{2app} \frac{(1 - v_{01}v_{02}/C^2)(1 - v_{02}V/C^2) + \dfrac{(v_{02} - v_{01})(V - v_{02})}{C^2}}{(1 - v_{02}V/C^2)\sqrt{(1 - v_{01}^2/C^2 - v_{02}^2/C^2 + v_{01}^2v_{02}^2/C^4}}$$

$$= T_{2app} \frac{(1 - v_{01}v_{02}/C^2) + \dfrac{(v_{02} - v_{01})}{C^2}\dfrac{(V - v_{02})}{1 - v_{02}V/C^2}}{\sqrt{1 - v_{01}^2/C^2 - v_{02}^2/C^2 + 2v_{01}v_{02}/C^2 + v_{01}^2v_{02}^2/C^4 - 2v_{01}v_{02}/C^2}}$$

$$= \frac{\dfrac{T_{2app}(1-v_{01}v_{02}/C^2) + \dfrac{(v_{02}-v_{01})}{C^2}X_{2app}}{1-v_{01}v_{02}/C^2}}{\sqrt{\dfrac{(1-v_{01}v_{02}/C^2)^2 - \dfrac{(v_{01}-v_{02})^2}{C^2}}{(1-v_{01}v_{02}/C^2)^2}}}.$$

Finally :

$$T_{1app} = \frac{T_{2app} + \dfrac{v_{02}-v_{01}}{1-v_{01}v_{02}/C^2}\dfrac{X_{2app}}{C^2}}{\sqrt{1 - \dfrac{1}{C^2}\left(\dfrac{v_{02}-v_{01}}{1-v_{01}v_{02}/C^2}\right)^2}}. \tag{27}$$

The same remarks as those concerning the space transformations can be made.

When $S_1$ is at rest in the Cosmic substratum, expression (27) reduces to:

$$T_0 = \frac{T_{2app} + v_{02}X_{2app}/C^2}{\sqrt{1-v_{02}^2/C^2}}.$$

And the reciprocal transformation takes the form:

$$T_{2app} = \frac{T_0 - v_{02}X_0/C^2}{\sqrt{1-v_{02}^2/C^2}}$$

a result which highlights the fact that the laws of nature are affected by the existence of the aether drift. (This result restricts the application of the relativity principle, and the time transformations (27) do not constitute a group in all generality.)

But with the apparent speeds $v_{12app}$ measured with contracted standards and with clocks slowed down by motion and synchronized with light signals, the time transformations take *the same mathematical form* as the conventional transformations between any pair of 'inertial frames', their general form being:

$$T_{1app} = \frac{T_{2app} + v_{12app}X_{2app}/C^2}{\sqrt{1-v_{12app}^2/C^2}} \text{ and } T_{2app} = \frac{T_{1app} - v_{12app}X_{1app}/C^2}{\sqrt{1-v_{12app}^2/C^2}}$$

and therefore the relativity principle *seems* to apply. Actually it does not really strictly apply when measurements are exact, and therefore it gives a distorted view of reality.

As we saw , $v_{12app}$ depends explicitly on the speeds $v_{01}$ and $v_{02}$, a fact which highlights the vital role of the aether in this derivation, while in conventional approaches the aether is hidden or nonexistent.